\documentclass[twocolumn,showpacs]{revtex4}
\usepackage[dvipdfm]{graphicx}
\usepackage{amsmath}
\makeatletter
\def\btt#1{\texttt{\@backslashchar#1}}%
\DeclareRobustCommand\bblash{\btt{\@backslashchar}}%
\makeatother

\begin{document}
\title{molecular dynamics investigation of dislocation pinning by a nanovoid in copper}
\author{Takahiro Hatano}
\affiliation{
Earthquake Research Institute, University of Tokyo, Tokyo, 113-0032 Japan}
\author{Hideki Matsui}
\affiliation{
Institute for Materials Research, Tohoku University, Sendai, 980-8577 Japan}
\date{\today}

\begin{abstract}
Interactions between an edge dislocation and a void in copper are investigated 
by means of molecular dynamics simulation.
The depinning stresses of the leading partial and of the trailing partial show qualitatively different behaviors.
The depinning stress of the trailing partial increases logarithmically with the void radius, 
while that of the leading partial shows a crossover at $1$ nm above which two partials are 
simultaneously trapped by the void.
The pinning angle, which characterizes the obstacle strength, approaches zero when the void radius exceeds $3$ nm.
No temperature dependence is found in the critical stress and the critical angle.
This is attributed to an absence of climb motion.
The distance between the void center and a glide plane asymmetrically affects the pinning strength.

\end{abstract}

\maketitle

\section{introduction}
\begin{figure}
\includegraphics[scale=0.4]{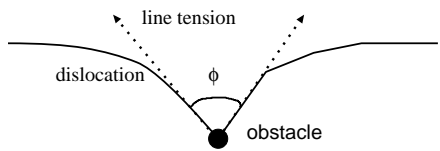}
\caption{A cusp formed at an obstacle. The angle $\phi$ between two tangential vectors 
is called the pinning angle.}
\label{cartoon}
\end{figure}

Voids are ubiquitous in irradiated metals and act as obstacles to dislocation motion as well as other radiation-induced defects:
e.g. stacking fault tetrahedra or helium bubbles.
Those obstacles results in the increase of yield stress and therefore 
play an important role in irradiation hardening.
To investigate the extent of hardening by those obstacles, there has been a model 
in which a dislocation is regarded as a continuous line with a constant line tension.
This is referred to as the uniform line tension model.
In the presence of obstacles, a dislocation is fixed to form a cusp at an obstacle.
The pinning angle $\phi$ is defined as the angle between two tangent vectors at a cusp.
(See FIG. \ref{cartoon}).
Then the restoring force to make a dislocation straight is written as $2\gamma\cos(\phi/2)$,
where $\gamma$ denotes the line tension.
We assume that a dislocation can penetrate an obstacle when the restoring force 
exceeds the critical value.
Since $\gamma$ is a constant, this condition is equivalent to $\phi\le\phi_c$, 
which we call the critical angle.
Note that stronger obstacles have smaller critical angles.
A dislocation bows out to form an arc between two obstacles 
until the pinning angle reaches its critical value.

For a periodic array of obstacles whose spacing is $L$, 
the critical resolved shear stress (CRSS) $\tau_c$ above which 
a dislocation can penetrate the array of obstacles is represented by 
\begin{equation}
\label{tauperiodic}
\tau_c=\frac{2\gamma}{bL}\cos\frac{\phi_c}{2}, 
\end{equation}
where $b$ denotes the Burgers vector length of the dislocation.
(The line tension $\gamma$ is given by the elastic theory 
and is often written as $Gb^2/2$, where $G$ represents the shear modulus.)

In more realistic situations, obstacles are distributed randomly on glide planes 
and the randomness plays a crucial role in dislocation dynamics.
In order to incorporate this effect, Foreman and Makin performed computer simulation 
of the dislocation motion on a glide plane with randomly distributed obstacles 
which have the same critical angle \cite{foreman}. 
They found that the dislocation propagation has two qualitatively different modes 
depending on the critical angle.
For obstacles of a small critical angle (i.e. strong obstacles), 
the dislocation propagation resembles dendritic growth \cite{cieplak}, 
while for a large critical angle (i.e. weak obstacles) 
the global form does not significantly deviate from the straight line.
Also $\tau_c$ are well described by \cite{note1}
\begin{equation}
\label{taurandom}
\tau_c=\left\{
\begin{array}{@{\,}ll}
\frac{2\gamma}{bL}\left(\cos\frac{\phi_c}{2}\right)^{\frac{3}{2}}, \ (\phi_c\ge\frac{5}{9}\pi) \\
\frac{1.6\gamma}{bL}\cos\frac{\phi_c}{2}, \ (\phi_c\le\frac{5}{9}\pi) 
\end{array}
\right.
\end{equation}
where $L$ denotes the inverse of the square root of the areal density of obstacles on a glide plane \cite{kocks}.

In this context, the critical angle $\phi_c$, which characterizes the obstacle strength, 
is an important parameter to discuss the extent of hardening.
However, estimation of the critical angle is not an easy task, 
because it involves the core structure of dislocations.
In this regard, extensive molecular dynamics (MD) simulations on the interaction 
between a dislocation and radiation-induced obstacles have been performed: 
e.g. stacking fault tetrahedra \cite{wirth}, interstitial Frank loops \cite{rodney}, 
voids and copper precipitates in bcc iron \cite{osetsky1,osetsky2,osetsky3}.

On the other hand, there are some experimental attempts to determine the critical angle 
utilizing transmission electron microscopy (TEM) \cite{robach,nogiwa}.
They seem to be promising but still should be complemented by MD simulations 
in terms of the spatial resolution. For example, subnanometer obstacles cannot be seen by TEM.

In this paper, along the line of the above computational and experimental studies, 
we wish to estimate the critical angle and the critical stress regarding voids in copper.
Especially, effects of the dissociation, temperature, and the distance from the void center to a glide plane 
are studied in detail. Note that we focus on an edge dislocation here.
Results on a screw dislocation will be presented elsewhere.

This paper is organized as follows.
In the section \ref{themodel}, we introduce the computational model; 
fcc copper including an edge dislocation and a void.
In the section \ref{behaviors}, the nature of the critical stress is investigated.
Especially, void size dependence and the temperature dependence is discussed.
The section \ref{thecriticalangle} deals with the critical angle, 
which is often measured by experiments (mainly TEM observation) to determine the obstacle strength.
The discussion enables us to compare the MD simulation with the experiments.
In the section \ref{impactparameterdependence}, effects of the distance 
between the void center and a glide plane are investigated.
In the section \ref{thecollapseprocess}, we discuss pinning strength of deformed voids 
in the context of dislocation channeling and plastic flow localization.
The last section \ref{discussionandconclusion} is devoted to discussions and conclusions.

\section{the model}
\label{themodel}

\subsection{the geometry}
We treat fcc copper in this paper.
As for the interatomic potential, we adopt the embedded-atom method 
of Finnis-Sinclair type \cite{finnis} and choose the parameters according to 
Ackland et al. \cite{ackland}. The lattice constant $a=3.615$ \AA.

\begin{figure}
\includegraphics[scale=0.4]{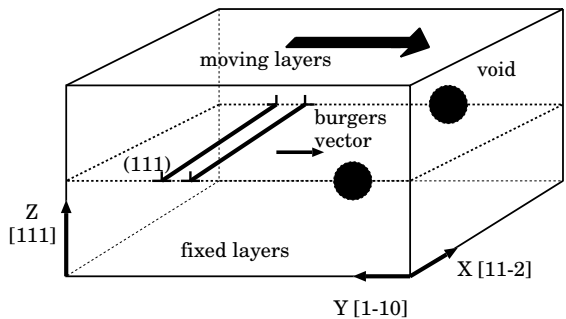}
\caption{Schematic of the system.
Periodic boundary conditions are assigned to the $x$ and the $y$ directions.
The system contains approximately $0.55$ million atoms.}
\label{configuration}
\end{figure}

The schematic of our system is shown in FIG. \ref{configuration}.
The $x$, $y$, and $z$ axes are taken as the $[11\bar{2}]$, $[1\bar{1}0]$, and $[111]$ directions, respectively.
The length of each dimension is $23$nm, $23$nm, and $15$ nm.
Periodic boundary conditions are employed in the $x$ and the $y$ directions.
That is, we consider dislocations of infinite length in the $x$ direction 
are periodically located in the $y$ direction.

Note that we have the surface only in the $z$ direction, both for $z>0$ and for $z<0$.
Following \cite{osetsky4}, three atomic layers of $[111]$ next to the lowest surface 
($z<0)$ are "the fixed layers" where velocities of the atoms always vanish.
Similarly, three atomic layers of $[111]$ next to the upper surface ($z>0)$ are "the moving layers" 
where velocities of the atoms are not given by the integration of force acting on them, 
but are given as a constant to cause the shear stress.
Namely, the strain rate is the control parameter: not the shear stress.
The moving surface moves to the $-y$ direction: i.e. $[\bar{1}10]$.

In order to introduce a void, atoms whose barycentric positions are 
in the spherical region $(x\pm L/2)^2+y^2+z^2<r^2$ are removed, 
where $r$ denotes the radius of the void.
To introduce an edge dislocation, atoms that belong to one $(1\bar{1}0)$ half plane ($z<0$) 
are removed and the rest of atoms are displaced by the strain field 
calculated from the elastic theory.
This procedure produces a perfect edge dislocation whose Burgers vector is $a/2[1\bar{1}0]$.
However, a perfect dislocation in an fcc crystal is energetically unstable to split 
into two partial dislocations whose Burgers vector length $b$ is $a/\sqrt{6}=1.48$ \AA.
\begin{equation}
\frac{a}{2}[\bar{1}10]\rightarrow \frac{a}{6}[\bar{2}11]+\frac{a}{6}[\bar{1}2\bar{1}].
\end{equation}
Since we wish to prepare two partial dislocations and a void in an initial system, 
atoms are suitably shifted from the original position by the steepest descent method 
in order to realize the dissociation.
In addition, since our system consists of the periodic array of dislocations 
due to the periodic boundary conditions in the $x$ direction, 
this procedure also incorporates the excess strain field caused by the next dislocations.
After certain time steps, the perfect dislocation dissociates to yield 
two partial dislocations separated by approximately $4$ nm.

Temperature is fixed to be $300$ K in this paper, except for the section \ref{temperaturedependence} 
where the temperature effect on the CRSS is investigated.
Velocities of atoms are given by random numbers which obey the Maxwell-Boltzmann distribution.
After a relatively short time required for phonon relaxation, 
"the moving layer", which is explained above, begins to displace to cause strain.

\subsection{the strain rate}
The strain rate $\dot{\epsilon}$ is an important parameter in MD simulations of dislocations.
In this paper, we set $\dot{\epsilon} = 8 \times 10^6$ [$\rm{sec}^{-1}$].
Although it seems still unrealistically fast deformation, 
the strain rate in MD simulation should not be directly compared with the macroscopic 
(or experimental) strain rate because the macroscopic strain rate involves only 
the average dislocation velocity.
Namely, both spatial and temporal fluctuations in dislocation velocity are neglected. 
The complete correspondence of microscopic dislocation velocity 
to the macroscopic strain rate is not clear at all 
unless we know the statistical property of space-time fluctuation in dislocation motion.

Calculation of the shear stress is noteworthy.
We define the shear stress as the forces in the $y$ direction 
acting on the unit area of the moving layers and the fixed layers.
However, this microscopic definition of the shear stress shows a large thermal fluctuation.
In addition, the inertial motion of the dislocation due to the high strain rate 
may enable depinning at lower stress \cite{osetsky1}.
In order to reduce these effects, simulation is performed twofold.
Namely, the representative point in the phase space 
(spanned by the positions and the momenta of all atoms) are recorded every $4.6$ ps.
We then take the point where depinning just begins 
and restart the simulation with $\dot{\epsilon}=0$.
In this relaxation process, the inertial effect is ruled out and 
the thermal fluctuation in the shear stress is averaged out.
Taking this relaxation process into account, 
the average strain rate becomes approximately $10^6$ [1/sec].

\section{Behaviors of the critical resolved shear stress}
\label{behaviors}

\subsection{temporal behavior}
\begin{figure}
\includegraphics[scale=0.8]{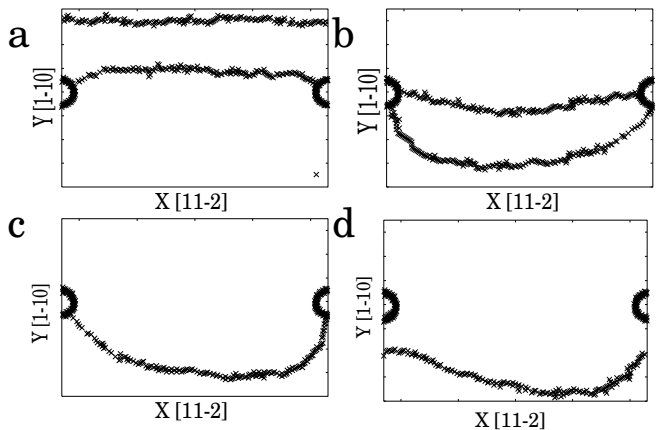}
\caption{Successive snapshots of a depinning process: 
(a) trapping of the leading partial, (b) just before the depinning of the leading partial, 
(c) just before depinning of the trailing partial, (d) depinning of the trailing partial. 
The void radius is $1$ nm. To visualize the void and the dislocations, 
atoms that have $12$ nearest neighbors are omitted.
(Atoms which form dislocations have $11$ or $13$ nearest neighbors, 
and the number of nearest neighbors of void surface atoms is less than $12$.)}
\label{snaps}
\end{figure}

First, we track the time evolution.
Under the shear stress caused by the boundary condition, two partial dislocations glide towards 
the $[\bar{1}10]$ direction: the $-y$ direction.
The Burgers vectors of the leading and the trailing partials are 
$[\bar{2}11]$ and $[\bar{1}2\bar{1}]$, respectively.
Snapshots of a pinning-depinning process are shown in FIG. \ref{snaps}.

\begin{figure}
\includegraphics[scale=0.38]{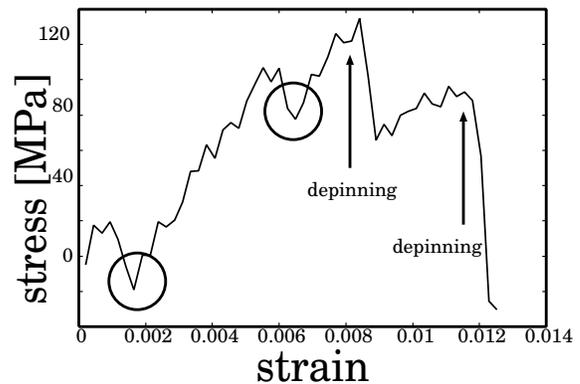}
\caption{Stress-strain relation. Two arrows indicate the depinning points of two partials.
Two circles which indicate sudden stress drops result from the attractive interaction 
between the dislocations and the void.}
\label{force}
\end{figure}

Note that there are two depinning processes of the leading and the trailing partials.
In the stress-strain relation (FIG. \ref{force}), we can see two peaks which correspond 
to the each depinning process.
We remark that the critical stress for the leading partial is always larger than that of the trailing partial.
This is because the leading partial has already escaped from the void and keeps gliding 
while the trailing partial is pinned by the void.
Then the width of the stacking fault ribbon extends to yield attractive force between the partials.
Namely, depinning of the trailing partial is assisted by the attractive force from the leading partial.
Also, note that the movement of the leading partial leads to the stress relaxation, 
which appears in the change of modulus in the stress-strain curve in FIG. \ref{force}.
That is, the modulus is $30$ GPa when the leading partial is pinned 
($0.002< \epsilon <0.006$), whereas the one after the leading partial is unpinned 
($0.008< \epsilon <0.012$) is $13$ GPa.

\subsection{void size dependence}
\begin{figure}
\includegraphics[scale=0.38]{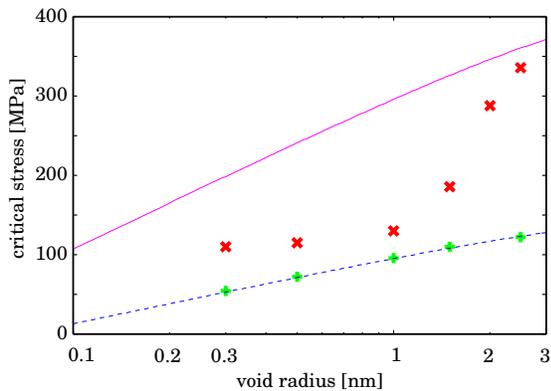}
\caption{Void radius dependence of the depinning stress for the leading partial ($\times$) 
and for the trailing partial ($+$). The dashed line denotes Eq. (\ref{tau_trailing}) 
with $T=37$ MPa and $B=0.14$ nm.
The dotted line denotes Eq. (\ref{sb}), which is the result from a continuous model calculation \cite{scattergood}.}
\label{r-CRSS}
\end{figure}

Then the critical stress is calculated for various voids of different radii (from $0.3$ to $2.5$ nm).
We measure the depinning stresses both for the leading partial and for the trailing partial.
An interesting feature arises from the comparison of the both stresses, which are shown in FIG. \ref{r-CRSS}.
We can see that they have different tendencies with respect to the void radius $r$.
The depinning stress of the trailing partial shows the well-known logarithmic dependence, 
while that of the leading partial (i.e. the yield stress) shows a crossover around $r\simeq 1$ nm.
We will discuss the difference more quantitatively in this subsection.

The depinning stress of the trailing partial can be described by the following relation.
\begin{equation}
\label{tau_trailing}
\tau_{\rm c} = T\log\left[\frac{2r}{B}(1+\frac{2r}{L})^{-1}\right],
\end{equation}
where $T$ and $B$ denote arbitrary constants.
The best fit is realized by letting $T=37$ MPa and $B=0.14$ nm.
Note that this logarithmic dependence has also been found in a continuous model by Scattergood and Bacon 
\cite{scattergood}, and also in the context of the Orowan mechanism \cite{bacon}.
Since their computation deals with only the critical stresses for pure edge dislocations 
and screw dislocations, it does not quantitatively apply to the partial dislocations.
At least, it can describe the qualitative behavior regarding the trailing partial.

On the other hand, the crossover of the yield stress cannot be explained in the above context.
It seems to be describable by piecewise logarithmic behavior; 
\begin{equation}
\label{tau_leading}
\tau_{\rm c} =\left\{
\begin{array}{@{\,}ll}
18 \log\left[\frac{2r}{1.5\times 10^{-3}}(1+\frac{2r}{L})^{-1}\right],\ \ (r\le 1)\\
280\log\left[\frac{2r}{1.2}(1+\frac{2r}{L})^{-1}\right].\ \ (r\ge 1)
\end{array}
\right.
\end{equation}
Note that the extent of hardening becomes much greater when the void radius exceeds $1$ nm. 
Although we cannot show the definite explanation of this phenomenon, 
a plausible reason lies in the interaction between two partials.
When the void radius is larger than $1$ nm, the two partials are simultaneously trapped by the void.
Then the depinning of the leading partial may be significantly affected by the trailing partial, 
since the distance between them is very close near the void (a few Burgers vectors).
Note that, for $r=0.3$ nm and $0.5$ nm, depinning of the leading partial takes place 
before the trapping of the trailing partial; i.e. the depinning of the leading partial 
is less affected by the trailing partial.
We also remark that a similar crossover due to the interaction between dislocations is observed 
in the critical stress of dislocation nucleation on the void surface \cite{hatano}.

Regardless of the void radius, the depinning of the trailing partial takes place 
when the leading partial is completely escaped from the void.
Although the trailing partial interacts with the leading partial via stacking fault ribbon 
and the elastic strain, it is not as strong as the one in the depinning process of the leading partial.
Therefore, it is reasonable that the unpinning stress for the trailing partial can be 
described in the framework of \cite{scattergood} where a single dislocation involves.

On the other hand, the yield stress for an edge dislocation should be interpreted as the maximum stress 
during the whole pinning process including two partials: that is, the critical stress for the leading partial.
It is interesting to compare the yield stress in the present simulation and the one in \cite{scattergood} 
in which the dissociation is not taken into account.
The yield stress estimated by Scattergood and Bacon reads 
\begin{equation}
\label{sb}
\tau_{\rm SB}=\frac{Gb}{2\pi L}\log\left[\frac{2r}{0.22b}(1+\frac{2r}{L})^{-1}\right], 
\end{equation}
which is also plotted in FIG. \ref{r-CRSS}.
We can see that Eq. (\ref{sb}) considerably overestimates the yield stress where $r\le 2.5$ nm.
Since it does apply to bcc iron \cite{osetsky1,osetsky2,osetsky3}, 
we can conclude that the dissociation makes depinning easier.

\subsection{temperature dependence}
\label{temperaturedependence}
\begin{figure}
\includegraphics[scale=0.35]{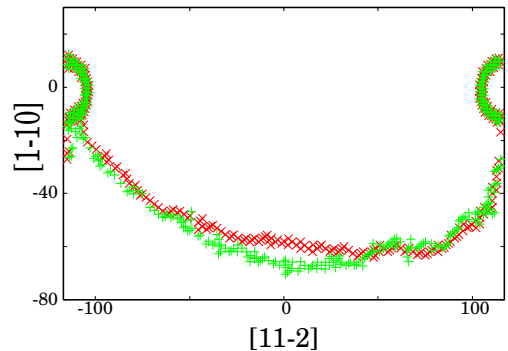}
\caption{Snapshots of dislocations (the trailing partials) just before depinning.
The trailing partial dislocation at $100$ K is represented by $\times$, 
and the one at $500$ K is $+$.
The subtle difference in the middle lies in the range of thermal fluctuation at $500$ K.}
\label{Tdependence}
\end{figure}

In this section we investigate temperature dependence of the critical stress.
We have calculated the critical stress for four different temperatures: 
$100$ K, $200$ K, $400$ K, and $500$ K.
However, we cannot find any differences regarding the critical stress between these calculations.
In FIG. \ref{Tdependence} we show the snapshots of dislocations just before depinning 
where the temperatures are $100$ K and $500$ K, respectively.
We can see no difference in the dislocation shape.
Also, the critical stress is almost the same: $130\pm 12$ MPa for $r=1$ nm and 
$288\pm 30$ MPa for $r=2.5$ nm.
Namely, temperature plays no role in depinning processes.

Note that this result is opposite to the simulation on bcc iron with a copper precipitate \cite{osetsky3}.
There, definite temperature dependence was observed from $0$ K to $500$ K.
The reason for the difference lies in the dissociation nature of dislocation.
In bcc iron where a dislocation does not dissociate, a perfect dislocation absorbs vacancies from a void 
(or the precipitate surface) and undergoes climb motion.
The climb motion is remarkable for larger voids 
where temperature dependence of the critical depinning stress is observed.
Note that smaller voids show less temperature dependence and the climb motion is weak there.
It implies that the climb motion is essential to the temperature dependence.
On the other hand, no climb motion is observed in the present simulation.
It is known that the climb motion is difficult in fcc crystals since dislocations dissociate.
Therefore, no temperature dependence is observed in the present simulation on fcc copper.

Another possible reason lies in the activation energy for depinning.
Although the precise estimation of the activation energy is difficult, 
it is at least larger than the energy of a dislocation whose length is equivalent to the void diameter.
(We neglect the step formation energy on the void surface.)
Dislocation energy is calculated by the line tension multiplied by the length.
The effective line tension is estimated to be $0.4$ nN by Eq. (\ref{estimategamma}).
(Please see the next section for the detail.)
Therefore, for the void of $2.0$ nm radius, the activation energy is 
approximately estimated as $1.6\times10^{-18}$ J.
It is equivalent to $400$ $k_BT$, where $k_B$ denotes the Boltzmann constant and $T=300$ K.
Since this is much larger than the thermal energy of involved atoms on the void surface 
(less than a hundred), it is plausible that thermal fluctuations cannot assist dislocation depinning.
In addition, please recall that the step formation energy is neglected and the actual value may be larger than that.
However, in the case of precipitates, the number of involving atoms is larger than in voids.
That may be another reason for the difference between the present simulation 
and the one of \cite{osetsky3}.

\section{The dislocation shape and the critical angle}
\label{thecriticalangle}
\subsection{an orientation dependent line tension model}
\begin{figure}
\includegraphics[scale=0.65]{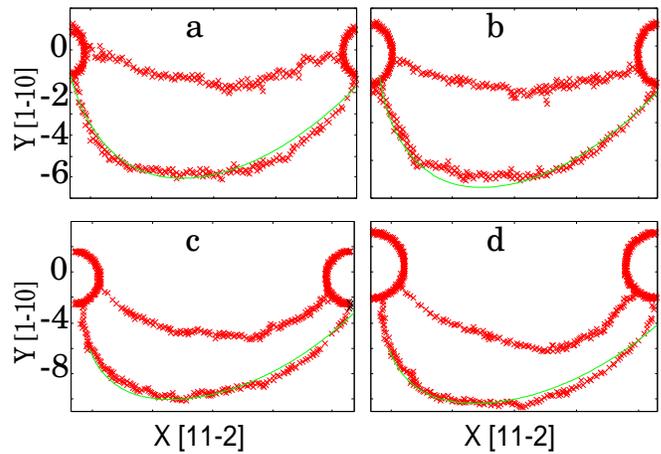}
\caption{Bowing dislocations obtained by the simulation for various voids 
(a: $r=1.0$ nm, b: $r=1.5$ nm, c: $r=2.0$ nm, d: $r=2.5$ nm.) 
The solid lines represent Eqs. (\ref{elliptic1}) and (\ref{elliptic2}), 
which are derived from an orientation dependent line tension model.
Note that the trailing partials also deform to a certain degree as the void becomes larger.}
\label{odfitting}
\end{figure}

As can be seen in FIG. \ref{snaps}, the bowing dislocation is asymmetric with respect to $x=0$.
Since the uniform line tension model predicts the symmetric form (an arc), we have to incorporate 
an orientation dependent line tension in order to explain the asymmetric dislocation shape.
Indeed, de Wit and Koehler \cite{dewit} obtained a solution for the dislocation shape.
\begin{eqnarray}
\label{elliptic1}
y&=& C_1 + \frac{1}{\sigma b}\left[E(\theta)\cos\delta -\frac{dE}{d\theta}\sin\delta\right], \\
\label{elliptic2}
x&=& C_2 + \frac{1}{\sigma b}\left[E(\theta)\sin\delta +\frac{dE}{d\theta}\cos\delta\right], 
\end{eqnarray}
where $\delta$ denotes the angle between the tangent line of a dislocation and the $x$ axis.
An orientation dependent line tension is denoted by $E(\theta)$, in which $\theta$ represents 
the angle between the tangent line of the dislocation and the Burgers vector.
If the concrete form of $E(\theta)$ is given, these equations can be numerically solved 
with an appropriate choice of the arbitrary constants $C_1$ and $C_2$.
Here, $E(\theta)$ is given as \cite{dewit} 
\begin{eqnarray}
\label{odlinetension1}
E(\theta)&=& \frac{b^2}{4\pi}\log\left(\frac{R}{r_0}\right) f(\theta),\\
\label{odlinetension2}
f(\theta) &\simeq& 59.3-16\cos2\theta-0.8\cos4\theta,\ \ [{\rm GPa}]
\end{eqnarray}
where $r_0$ is the (arbitrary) core cutoff length.
In FIG. \ref{odfitting}, Eqs. (\ref{odlinetension1}) and (\ref{odlinetension2}) are fitted with the simulation result.
We can see that the fittings are quite satisfactory.

\subsection{determination of the critical angle}
In macroscopic materials, random configuration of obstacles plays an important role in the dislocation dynamics, 
as was discussed in the section I. In order to apply Eq. (\ref{taurandom}) to a practical situation, 
we wish to estimate the critical angle from the simulation.
Note that the critical angle $\phi_c$ is defined by the angle between two tangent lines 
of the dislocation at the void surface.
Because the critical angles are slightly different for two partial dislocations, we measure the both.

\begin{figure}
\includegraphics[scale=0.4]{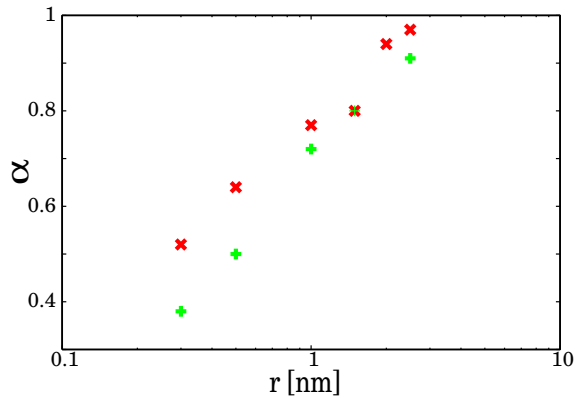}
\caption{Void radius dependence of $\alpha$ for the leading partial ($\times$) 
and for the trailing partial ($+$).}
\label{r-alpha}
\end{figure}

The pinning strength $\alpha =\cos(\phi_c/2)$ with respect to the void radius is shown in FIG. \ref{r-alpha}.
We remark that the pinning strength $\alpha$ again obeys a logarithmic law.
\begin{equation}
\label{fittingalpha}
\alpha =A \log\frac{2r}{B(1+\frac{2r}{L})}.
\end{equation}
The constants are $A=0.24$ and $B=0.07$ for the leading partial, and $A=0.28$ and $B=0.15$ for the trailing partial.
By extrapolation, $\alpha$ reaches $1$ when the void radius exceeds $3$ nm, 
as shown in FIG. \ref{r-alpha}.
Note that the tendency has also been observed in the previous simulations on bcc iron 
\cite{osetsky1,osetsky2,osetsky3}.
\begin{figure}
\includegraphics[scale=0.4]{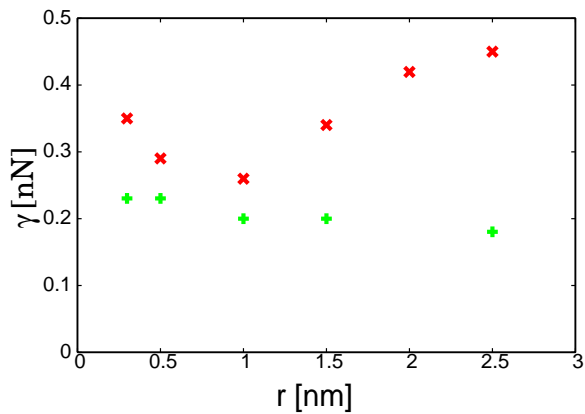}
\caption{Effective line tensions estimated via Eq. (\ref{estimategamma}).
Those of the leading partial and of the trailing partial are represented by $\times$ and $+$, respectively.
Note that the void radius dependence results from the difference in the configuration of dislocations: 
the degree of bowing and the interactions of two partials.}
\label{r-gamma}
\end{figure}

Meanwhile, it should be remarked that a crossover is not observed in the pinning strength $\alpha$.
From Eq. (\ref{tauperiodic}), the crossover of $\tau_c$ should be attributed to that of $\gamma$, 
which can be written as 
\begin{equation}
\label{estimategamma}
\gamma=\frac{b\tau_cL}{2\alpha}.
\end{equation}
The calculated line tensions are shown in FIG. \ref{r-gamma}, in which the line tension of the leading partial 
shows the minimum at $r=1.0$ nm while that of the trailing partial monotonically decreases.
The explanation for the decrease lies in the deformations (i.e. bowouts) of dislocations.
As the void gets larger from $r=0$, the extent of bowout becomes greater.
This results in the prevalence of the screw component, since the partial dislocation is initially edge-like.
The deformation lowers the effective line tension: i.e. the energy per unit length.
(However, the total energy increases due to the elongation.)
The above mechanism also explains the decrease of the line tension of the leading partial for $r\le 1$ nm.

On the other hand, the increase of the line tension of the leading partial for $r\ge 1$ nm 
results from the interaction with the trailing partial. 
For larger voids, the trailing partial is also pinned by the void and bends 
during the depinning process of the leading partial, as can be seen in FIG. \ref{odfitting}.
That effectively increases the line tension of the leading partial.


\section{effects of the impact parameter}
\label{impactparameterdependence}
\begin{figure}
\includegraphics[scale=0.5]{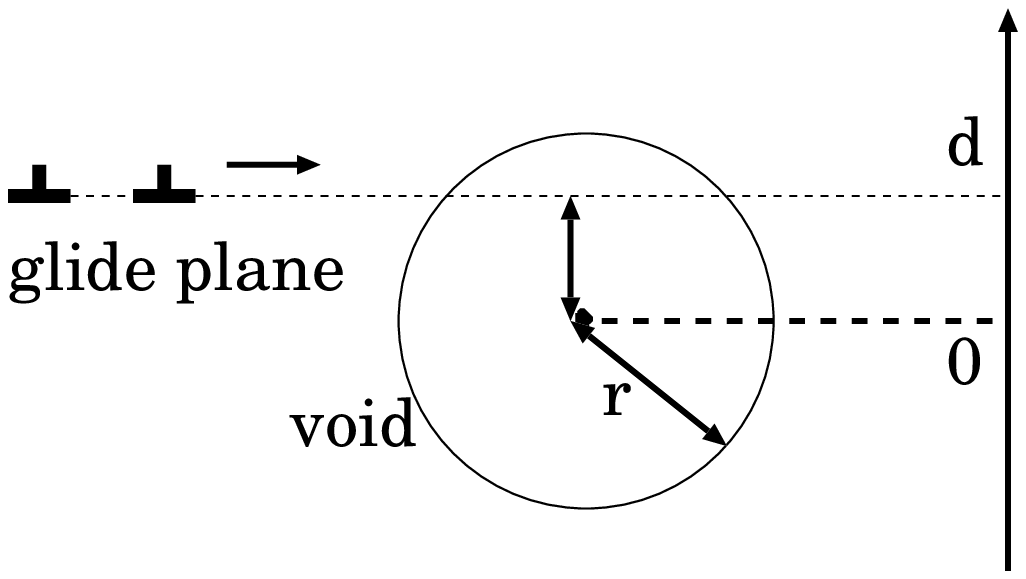}
\caption{The impact parameter $d$ is defined by the distance from the void center to the glide plane.
Note that the lower part of the void corresponds to the negative values of $d$.} 
\label{impactparameter}
\end{figure}

So far, we have limited ourselves to the situation where a dislocation penetrates the void center.
This is rather a special case, because the relative position of a void to a glide plane may be arbitrary.
In this section, we change the distance between the void center and a glide plane.
We call the distance "the impact parameter", which is denoted by $d$.
(See FIG. \ref{impactparameter}.)

\begin{figure}
\includegraphics[scale=1.1]{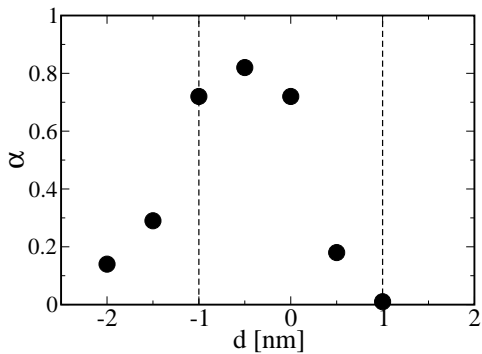}
\caption{Impact parameter dependence of the pinning strength $\alpha$ for the trailing partial.
Void radius is $1.0$ nm (indicated by the dashed lines).}
\label{d-alpha}
\end{figure}

The pinning strength $\alpha$ is determined for the various impact parameters.
The result is shown in FIG. \ref{d-alpha}, which shows asymmetric dependence of $\alpha$ on $d$.
Note that the dislocation is pinned even when it is not in contact with the void.
It implies that, as well as the core energy, the elastic strain around the dislocation 
plays an important role in pinning processes.
In addition, the asymmetry regarding $d=0$ comes from the nature of strain field 
around an edge dislocation: i.e. existence of hydrostatic pressure caused by the extra atomic plane.
Especially, the fact that pinning strength for $d>0$ becomes considerably weak 
suggests that the hydrostatic pressure is dominant over the shear stress.

Also, we remark that strong pinning ($\alpha\ge 0.5$) occurs only where $-1.0r\le d\le0$: 
i.e. the lower half of the void.
This area accounts for approximately $30$ or $40$ percent of the whole pinning region, 
while the rest involves relatively weak pinning.
The large variance of the pinning strength distribution for a single void suggests 
the reconsideration of the same pinning strength assumption in dislocation dynamics simulations.
In the section \ref{discussionandconclusion}, we discuss how to incorporate this effect 
into the estimation of the CRSS in the framework of Eq. (\ref{taurandom}).

\section{effects of void deformation after the passage of several dislocations}
\label{thecollapseprocess}
When a void is sheared by a dislocation, two parts which are divided by the glide plane 
are displaced to each other by the Burgers vector.
After the passages of several dislocations, it may collapse and lose the pinning ability.
For example, collapse of stacking fault tetrahedra by the passage of dislocations 
is both experimentally \cite{matsukawa} and computationally \cite{wirth} observed.
This phenomenon is believed to be responsible for the formation of dislocation channel 
and the localization of plastic flow, which recently invokes attention 
including some computational studies \cite{delarubia}.
In this section, effects of the void deformation on pinning strength are discussed 
based on the motivation described above.

We remark that another possible mechanism of the void deformation is vacancy absorption by 
(and the climb motion of) an edge dislocation, as was discussed in the section \ref{temperaturedependence}.
However, no climb motion was seen in our simulations, because the climb is difficult in fcc metals 
due to its dissociation.
Hence, we do not consider the void contraction by the vacancy emission to dislocations.
We concentrate on the effect of the relative deformation with respect to a glide plane.

\begin{figure}
\includegraphics[scale=0.6]{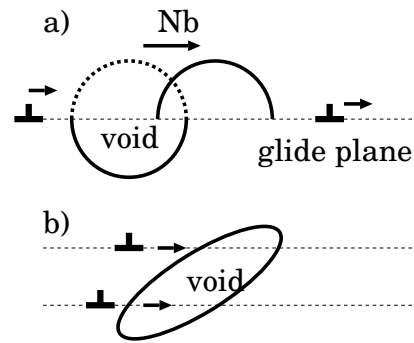}
\caption{Schematic of the void deformation by the passages of $N$ edge dislocations: 
a) There is a single glide plane which cuts the void center.
b) Glide planes are uniformly distributed. Each planes has $N$ dislocations 
which are to penetrate the void.}
\label{deformation}
\end{figure}

We prepare the deformed void as shown in FIG. \ref{deformation}.
First we prepare a spherical void.
Then, instead of iterating the pinning simulations, atoms located above the glide plane 
are displaced by the Burgers vector $a/2[\bar{1}10]$.
Iterating this procedure for $N$ times is equivalent to the passage of $N$ edge dislocations.
We set two configurations.
In the case (a), the glide plane on which the dislocations move is assumed to be at the void center, 
whereas glide planes are uniformly distributed in the case (b).

In the case (a), we set $N=5$ and $N=10$.
Namely, the void is assumed to be penetrated by dislocations on the same glide plane $5$ times or $10$ times.
The pinning strength $\alpha$ is found to be $0.7$ for the both cases, 
which is almost the same value with the one for the spherical void.
This is consistent with the result obtained in the last section that the upper part of the void is dominant 
in pinning of edge dislocations because of the hydrostatic pressure.
In the case (b), we test $N=2$ and cannot find any difference from the non-deformed case regarding the critical angle.
Thus, as far as the vacancy absorption mechanism (i.e. climb) is absent, 
the pinning strength is not seriously altered by the passage of dislocations. 

\section{discussions and concluding remarks}
\label{discussionandconclusion}

\subsection{practical applications}
Let us estimate the yield stress of copper in which voids are randomly distributed.
First, we determine the average spacing $L$ between obstacles on a glide plane.
The areal density is represented as $2r\rho$, where $\rho$ denotes the number density of voids per unit volume.
Then the average spacing on a glide plane is written as $L=1/\sqrt{2r\rho}$.
Since the expression includes all voids which intersects a glide plane, 
their impact parameters are randomly distributed from $-r$ to $r$.
Since we have seen that the pinning strength $\alpha$ considerably changes with the impact parameter $d$ 
in the section \ref{impactparameterdependence}, we wish to incorporate this result.
However, there is no simulation which considers this effect at this point.
Therefore, we have to resort a rough approximation here. 

From FIG. \ref{d-alpha}, strong pinning ($\alpha\ge 0.4$) occurs only where $-1.2r\le d \le 0.2r$.
We neglect the rest. Namely, it is assumed that only this region is responsible for pinning.
We use $1.4r$ instead of the diameter $2r$; then $L\simeq 1/\sqrt{1.4r\rho}$.
We can rewrite Eq. (\ref{taurandom}) as  
\begin{equation}
\tau_c=\frac{1.8\gamma\alpha}{b}\sqrt{1.4r\rho}, 
\end{equation}
where the line tension $\gamma$ should be interpreted as the effective one 
that is determined by the present simulation, FIG. \ref{r-gamma}. 

For example, an irradiated copper specimen includes voids whose average diameter $2r=4.1$ nm, 
and the number density $\rho$ is $2.9\times10^{22}$ $\rm{m}^{-3}$ \cite{nita}.
From the present simulation, the line tension $\gamma$ and the pinning strength $\alpha$ are 
estimated as $0.42$ nN and $0.9$, respectively.
The effective areal density is then calculated as $8.4\times10^{-5}$ $\rm{nm}^{-2}$ 
(i.e. the average spacing is $110$ nm).
This yields $\tau_c=40$ MPa, which is quite a reasonable value.
In order to give more precise predictions, Eq. (\ref{taurandom}) should be modified to include 
the effects of the impact parameter and the orientation dependent line tension.

\subsection{comparison with a continuous model with self-interaction}
Scattergood et al. have presented the calculation on the dislocation pinning by a void 
based on the framework of Bacon et al. \cite{bacon}.
It is a continuous model that incorporates the self-interaction of a dislocation.
Since their system also consists of the periodic array of voids, we wish to compare their result to ours.

Let us briefly review their discussion.
They have speculated that depinning concerns the line tension near the pinning point, where a dislocation forms a dipole.
Since a dislocation rotates by almost $\pi/2$ there, we regard the effective line tension as that of its counterpart.
For example, as for an edge dislocation, the effective line tension upon depinning is that of a screw dislocation.
Then the depinning stress for an edge dislocation is expressed by 
\begin{eqnarray}
\label{bks1}
\tau&=&\frac{2\gamma_{\rm{eff}}}{Lb}\\
\label{bks2}
\gamma_{\rm{eff}}&=&\frac{Gb^2}{4\pi}\log\bar{R},\\
\label{bks3}
\frac{1}{\bar{R}}&=&\frac{B}{L-2r}+\frac{B}{2r},
\end{eqnarray}
where $\bar{R}$ should be interpreted as the effective outer cutoff divided by the core cutoff.
Note that $B$ itself is not the core cutoff but an unknown function of the core cutoff.
The above equations yield Eq. (\ref{sb}).

Their discussion can be extended to dislocations of arbitrary orientation.
\begin{equation}
\label{bks4}
\gamma_{\rm eff}=\frac{Gb^2}{4\pi(1-\nu)}\left[1-\nu\cos^2(\frac{\pi}{2}-\theta)\right]\log{\bar R},
\end{equation}
where $\nu$ and $\theta$ denote Poisson's ratio and the angle between the dislocation line 
and the Burgers vector, respectively.
In the present case, substituting $\theta=\pi/3$ (for the partials) into Eq. (\ref{bks4}) 
yields $\tau=30\log{\bar R}$ [MPa].
Recall that Eq. (\ref{bks4}) is an expression for the resolved shear stress with respect to 
the Burgers vector of a partial dislocation.
It is equivalent to $\tau=35\log{\bar R}$ [MPa] with respect to the Burgers vector of an original perfect dislocation.
Indeed, it shows an excellent agreement with the critical stress of the trailing partial.

However, the problem is the arbitrary parameter $B$.
This arbitrariness may vanish if we can calculate the deformed (curved) effect on the dislocation energy, 
although it is a very complicated problem.
At least, we confirm that the extended form of the proportional constant Eq. (\ref{bks4}) agrees with 
the present simulation, if there is no strong interaction between the partials.

\subsection{conclusion}
We calculated the critical stress and the critical pinning angle for the interaction 
between an edge dislocation and a void in fcc copper.
Dissociation of a dislocation plays an important role in the behaviors of the critical stress:
i) It is much lower than the estimation of Scattergood and Bacon, which does not consider dissociation.
ii) It suddenly increases at a certain void radius where two partials can be simultaneously trapped.
iii) The depinning stress of the trailing partial does agree with that of Scattergood and Bacon, 
since the leading partial moves far from the void.

We also found that there is no temperature dependence in the critical stress and the pinning angle.
This is opposite to the previous simulation on bcc iron \cite{osetsky1,osetsky2,osetsky3}.
The difference comes from the presence (in bcc) or the absence (in fcc) of climb motion.

The pinning strength $\cos(\phi_c/2)$ obeys the empirical logarithmic law 
which has been found in \cite{scattergood,bacon}.
The distance between the void center and the glide plane (the impact parameter) 
is found to affect the pinning strength in asymmetric manner.
This is due to the hydrostatic pressure around an edge dislocation.
Hence, it is interesting to compare the result with that of a screw dislocation, 
which is a work in progress.

The impact parameter dependence of the critical angle also suggests 
the importance of randomness in the pinning strength.
Even if the system contains voids of the same radius, the cross section on a glide plane is randomly distributed.
Hence we have to incorporate the randomness in the pinning strength.
It is not straightforward to deduce this effect from the existing simulations 
which treats only two kinds of obstacles \cite{foreman2}.
Investigation of a continuous model with random pinning angles will be interesting 
to see how the impact parameter dependence affects the macroscopic dislocation motion.

\acknowledgements
The authors gratefully acknowledge Nobuyasu Nita for useful discussions regarding experimental situations.
They also thank Yuhki Satoh and Hideo Kaburaki for discussions and valuable comments.


\begin{thebibliography}{99}
\bibitem{note1}
The validity of the definition of critical resolved shear stress in the dendritic-growth mode is 
questionable because of the considerable inhomogeneity of plastic flow.
In addition, the self-interaction can no longer be negligible there.
\end{thebibliography}
\end{document}